\documentclass[10pt,superscriptaddress,nofootinbib,notitlepage,tightenlines,onecolumn, pra]{revtex4-1}

\usepackage{amsthm,amsmath,amssymb}

\usepackage{mathrsfs}
\usepackage{amsthm}
\usepackage{amsmath}
\usepackage{amssymb}
\usepackage{graphicx}
\usepackage{epstopdf}
\usepackage{url}
\usepackage{float}
\usepackage{bm}
\usepackage{ifthen}
\usepackage[usenames,dvipsnames]{color}
\usepackage{mathrsfs}
\usepackage[colorlinks=true,citecolor=blue,urlcolor=black]{hyperref}
\usepackage{float}

\newcommand{\be}{\begin{equation}}
\newcommand{\ee}{\end{equation}}

\newcommand{\sket}[1]{{\ensuremath{\lvert#1\rangle}}}
\newcommand{\lket}[1]{{\ensuremath{\left\lvert#1\right\rangle}}}
\newcommand{\ket}[1]{\if@display\lket{#1}\else\sket{#1}\fi}

\newcommand{\sbra}[1]{{\ensuremath{\langle#1\rvert}}}
\newcommand{\lbra}[1]{{\ensuremath{\left\langle#1\right\rvert}}}
\newcommand{\bra}[1]{\if@display\lbra{#1}\else\sbra{#1}\fi}

\newcommand{\sbraket}[2]{{\ensuremath{\langle#1\rvert#2\rangle}}}
\newcommand{\lbraket}[2]{{\ensuremath{\left\langle#1\!\left\rvert\vphantom{#1}#2\right.\!\right\rangle}}}
\newcommand{\braket}[2]{\if@display\lbraket{#1}{#2}\else\sbraket{#1}{#2}\fi}

\newcommand{\sketbra}[2]{{\ensuremath{\lvert #1\rangle\!\langle #2\rvert}}}
\newcommand{\lketbra}[2]{{\ensuremath{\left\lvert #1\right\rangle\!\!\left\langle #2\right\rvert}}}
\newcommand{\ketbra}[2]{\if@display\lketbra{#1}{#2}\else\sketbra{#1}{#2}\fi}

\theoremstyle{plain}

\theoremstyle{definition}

\begin{document}

\title{Secure and efficient synchronization scheme for quantum key distribution}

\author{Peng Liu}
\affiliation{National Laboratory of Solid State Microstructures and School of Physics, Nanjing University, Nanjing 210093, China}
\affiliation{Zhongchuangwei Quantum Co., Ltd., Beijing 101400, China}

\author{Hua-Lei Yin}\email{hlyin@nju.edu.cn}
\affiliation{National Laboratory of Solid State Microstructures and School of Physics, Nanjing University, Nanjing 210093, China}
\affiliation{Zhongchuangwei Quantum Co., Ltd., Beijing 101400, China}
\affiliation{Department of Physics and Zhejiang Institute of Modern Physics, Zhejiang University, Hangzhou 310027, China}


\begin{abstract}
To establish a time reference frame between two users in quantum key distribution, a synchronization calibration process is usually applied for the case of using gated mode single-photon detectors (SPDs). Traditionally, the synchronization calibration is independently implemented by the line length measurement for each SPD. However, this will leave a loophole which has been experimentally demonstrated by a special attack. Here, we propose an alternative synchronization scheme by fixing the relative delay of the signal time window among all SPDs and jointly performing the line length measurement with multiple SPDs under combining low-precision with high-precision synchronization. The new scheme is not only immune to the vulnerability but also improves the synchronization time from usually a few seconds to tens of milliseconds.
\end{abstract}

\maketitle
\section{Introduction}

Quantum key distribution (QKD)~\cite{bennett2014quantum,ekert1991quantum} is one of the most successful applications of quantum information science.
It promises unconditionally secure key generation between two distant parties via the fundamental laws of quantum physics~\cite{gisin2002quantum,scarani2009security}, rather than unproven computational complexity assumptions.
The best-known QKD protocol is the Bennett-Brassard-1984 (BB84) protocol~\cite{bennett2014quantum}, combining the decoy-state method~\cite{hwang2003quantum,wang2005beating,lo2005decoy}, which has been researched extensively both theoretically~\cite{shor2000simple,gottesman2004security,tomamichel2012tight,lim2014concise} and experimentally~\cite{rosenberg2007long,schmitt2007experimental,peng2007experimental,frohlich2013quantum,liao2017satellite}.
Currently, many practical BB84 QKD systems use optical communication fibers as quantum channels and work at the telecom wavelengths of around either 1550 nm or 1310 nm~\cite{yuan2009practical,sasaki2011field,wang2017long}.
The detection at these wavelengths is often performed by the single-photon detector (SPD) with InGaAs avalanche photodiodes operated in Geiger mode~\cite{hadfield2009single}, i.e., it is only activated for a narrow window when a signal pulse is expected to arrive.
Therefore, Alice and Bob should synchronize the optical pulse with the center of SPD's open window in the calibration routine.

Obviously, in this design, the synchronization between Alice's source and Bob's detectors is critical because the receiver will not identify the quantum signal correctly without the time gate synchronized to the photon arrival. How to choose an appropriate synchronization in the fiber network is worth considering carefully.
Improper synchronization will not only affect the efficiency (long synchronization time) but also compromise the security of the QKD system.
For example, the synchronization process is independently implemented by the line length measurement for each SPD~\cite{peng2007experimental,mink2008quantum}, i.e., only using those detected pulse count from one detector to align the time of that detector. However, this synchronization calibration process is not efficient. More importantly, it will leave a loophole for Eve to attack the QKD system. During the synchronization calibration phase, Eve can shift the arrival time of each synchronization pulse and cause the detector efficiency mismatch for gated mode SPD.
After the synchronization calibration process, Eve can use the fake-state attack~\cite{makarov2005faked,makarov2006effects,makarov2008fakes} or time-shift attack~\cite{qi2007time,zhao2008quantum} to successfully steal the secret key, which has been experimentally
demonstrated in a commercial plug-and-play phase-coding QKD system~\cite{jain2011device}.

In this paper, we present a secure and efficient synchronization method for QKD.
Based on the assumption of the fixed relative delay of signal time window among all SPDs, it is secure against the synchronization calibration attack~\cite{jain2011device}. Furthermore, the time of the synchronization calibration will be greatly decreased by using the pulse count from different detectors jointly.
This paper is organized as follows. In the second part, we first introduce a synchronization calibration attack for the traditional polarization-coding system, which independently implements the line length measurement for each SPD~\cite{peng2007experimental}. Then, we experimentally verify an assumption that the relative delay of signal time window among different SPDs in QKD system can be fixed even after a long time running. In the third and fourth parts, we propose and experimentally demonstrate a secure and efficient synchronization scheme. For each synchronization calibration process, the required time of the new scheme is only tens of milliseconds. In the final part, we have a conclusion.

\section{Synchronization loophole and countermeasure}
During the synchronization calibration process, Eve can replace the synchronization signal with his own signal and cause detector-efficiency mismatch if the line length measurement for each SPD is independently implemented~\cite{jain2011device}.
Taking a polarization-coding QKD system as an example, Eve can set the horizontal pulse $\vert H \rangle$ to arrive at $t_0$, and the vertical pulse $\vert V \rangle$ at $t_1$ in the synchronization calibration process.
At the receiver's side, the time window that has the largest number of triggering is considered as a signal time window for the detector~\cite{pljonkin2016single,rumyantsev2015synchronization,pljonkin2017features}.
Consequently, the time window for $\vert H \rangle$ is $t_0$ and the time window for $\vert V \rangle$ is $t_1$.
Moreover, Eve could set $t_0$ and $t_1$ carefully so that the detector efficiencies of the detectors differ greatly.
After the synchronization calibration process, Eve can use fake-state attack~\cite{makarov2005faked,makarov2006effects,makarov2008fakes} or time-shift attack~\cite{qi2007time,zhao2008quantum} to acquire the key.

\begin{figure}[h!]
\centering
  \includegraphics[width=12cm]{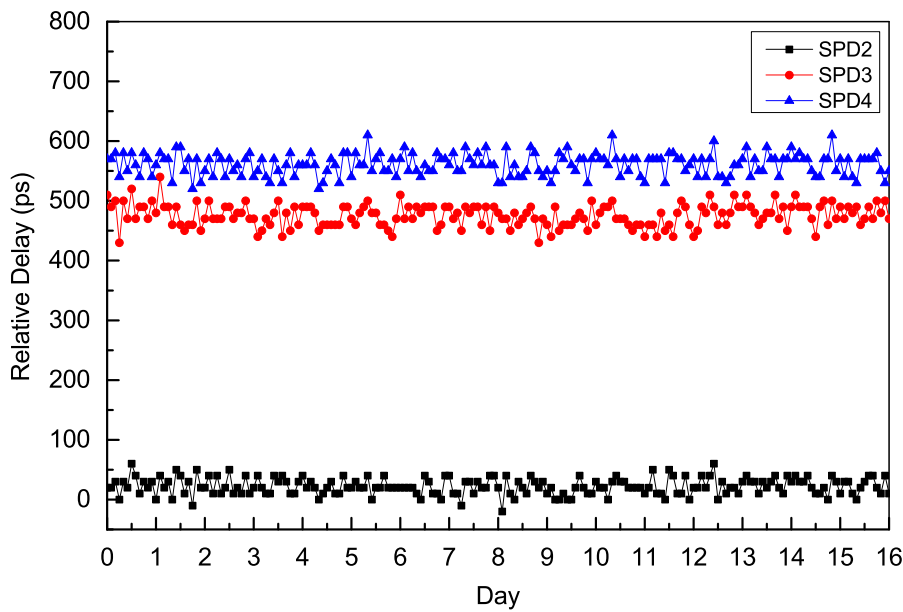}
\caption{The experimental result.}
\label{figure:The experimental result}
\end{figure}

In order to circumvent this loophole, we assume that the relative delay of signal time window among different SPDs can be fixed since the optical lengths and electrical delays for different detectors almost unchange at the receiver. We run one QKD system for $16$ days to record the relative delay of signal time window among different SPDs. The experimental result supports our hypothesis which can be found in Fig. ~\ref{figure:The experimental result}. We remark that the time-shift attack cannot be immune if one only uses our synchronization scheme, such as the case of the difference in detector efficiency characteristics.

\section{New synchronization scheme}
\begin{figure}[h!]
\centering
  \includegraphics[width=12cm]{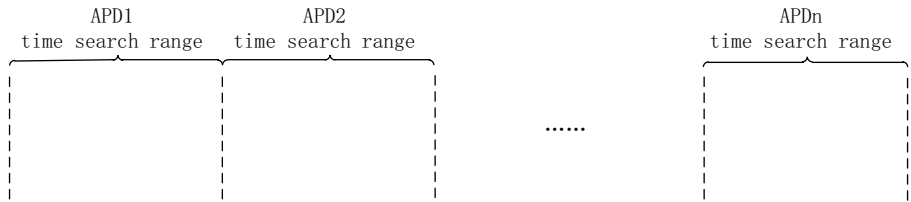}
\caption{Schematic diagram of parallel synchronization.}
\label{figure:1}
\end{figure}

Here, we propose a secure and efficient synchronization scheme. On the one hand, we exploit the fact that the difference between the signal time window of multiple detectors remains the same. On the other hand, we implement a parallel synchronization method, where the time search range is divided evenly into $N$ portions and each portion is searched by only one detector. The time search can be simultaneously performed by using multiple detectors.
Therefore, the time of the synchronization process can be considerably decreased. A schematic diagram of parallel synchronization is illustrated in Fig.\ref{figure:1}.

The repetition rate of a QKD system is typically from tens of megahertz to gigahertz, while synchronization accuracy typically reaches tens of picoseconds.
If the traversal method is utilized to finish the synchronization, the number of time windows that need to be searched can reach several hundreds to several thousands. Therefore, this synchronization method is extremely inefficient.

Here, we use a combination of low-precision and high-precision synchronization to reduce the number of searches.
First, a low-precision traversal search is utilized to confirm the approximate time range of the optical pulse over the full-time range.
The high-precision search is then used to obtain the signal time window in this relatively small time range.
The high-precision search scheme can be applied by the dichotomy method or other high-performance search algorithms.
By adopting the synchronization scheme combined with low-precision and high-precision, the search number can be significantly reduced, and the accuracy of the synchronization can be guaranteed.
First, we need to calibrate the synchronization time difference among different SPDs in the lab or some other safe zone.
In a BB84-QKD system, typically, Bob uses four separate SPDs, which are labeled as ${\rm SPD_{1}}$, ${\rm SPD_{2}}$, ${\rm SPD_{3}}$ and ${\rm SPD_{4}}$. Two SPDs are used for the X basis and two SPDs are used for the Z basis.
Based on the signal time window $T_1$ of ${\rm SPD_{1}}$, the differences between the signal time window of the remaining detectors ${\rm SPD_{\emph{i}}}$ and ${\rm SPD_{1}}$ are recorded as $\Delta T_{i}$, which can be given by
\begin{equation}
\Delta T_i = T_i - T_1,
\end{equation}
where $T_{i}$ is the signal time window of detector ${\rm SPD_{\emph{i}}}$, and $i=2,~3,~4$.

In the low-precision synchronization process, the receiver of the QKD system calculates the time search range of each detector according to the number of detectors ($N$), the relative delay value of each detector, and the repetition rate of the QKD system $f$.
The full time search range is divided evenly into $N$ portions.
For the $i$th detector, the time search range $Ra_i$ is:
\begin{equation}
Ra_i=[\frac{i-1}{Nf}+\Delta T_i,\frac{i}{Nf}+\Delta T_i]
\end{equation}
The requirement for low precision is no more than the full width at half maximum of the optical pulse. A low-precision synchronous search is finished when all detectors complete the search.

According to the detector count data of all the detectors in the low-precision process, we can find the time window $t_{l}$ and the detector number $n_{l}$ corresponding to the maximum detector count.
The time search range of high-precision synchronization is calculated based on $t_{l}$ and $n_{l}$.
The high-precision bit synchronization process and the low-precision bit synchronization process are roughly identical.
The difference is that, firstly, in order to ensure the accuracy of synchronization, the accumulation time of the detector count in the high-precision synchronization process is usually greater than that in the low-precision synchronization process.
Secondly, the high-precision synchronization process can further improve the efficiency by using the dichotomy algorithm.

According to the detector count data of all the detectors in the high-precision process, one need to find the time window $t_{h}$ and the detector number $n_{h}$ corresponding to the maximum detector count.
For the $m$th detector, the signal time window is calculated as follows: $t_{m}=t_{h}$ if $m=n_{h}$, otherwise $t_{m}=t_{h}+(\Delta T_m-\Delta T_{n_{h}})$.

\section{Experimental result}

\begin{figure}[h!]
\centering
  \includegraphics[width=12cm]{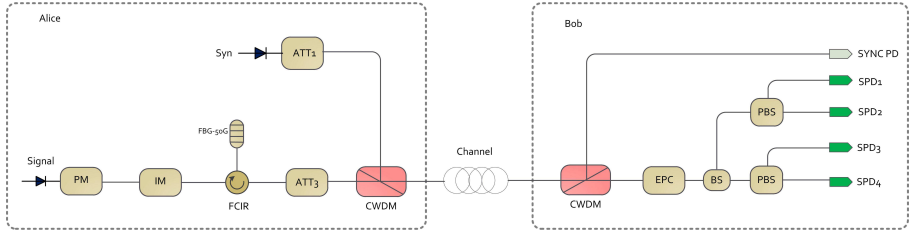}
\caption{Schematic diagram of polarization-coding QKD.
PM, polarization modulator.
IM, intensity modulator.
FBG-50G, fiber Bragg grating.
ATT, optical attenuator.
CWDM, coarse wavelength division multiplexing.
EPC, electronic polarization controller.
PBS, polarization beam splitter.
SPD, single-photon detector.
}
\label{figure:experiment setup}
\end{figure}

Fig.\ref{figure:experiment setup} displays a schematic diagram of our polarization-coding decoy-state QKD system setup, including transmitter (Alice), quantum channel and receiver (Bob).
At Alice's side, we employ the phase-randomized weak-coherent pulse of 500 ps duration emitted by a 1550.12nm distributed feedback laser operating at 100 MHz as a source. The randomly changed intensity of each pulse is realized by an intensity modulator.
The polarization of the signal is encoded by a fast polarization modulator.
Intensity modulation and polarization modulation are controlled by a true random number generator.
At Bob's side, we use an electronic polarization controller to recover the polarization alignment before his polarization decoding optics module.
Four InGaAs avalanche photodiodes operated in Geiger mode are used for single-photon detection at a clock rate of 100 MHz.
The detection efficiency is $15.3\%$ while the dark count probability is approximately $8.0 \times 10^{-7}$ per gate.
Between Alice and Bob, we use 50km standard single mode fiber SMF-28 with an attenuation loss of $10.3$ dB as the quantum channel.
During the synchronization calibration process, the intensity of each pulse is changed to 3 to accelerate the entire process.

\subsection{Method I}

\begin{figure}[h!]
\centering
  \includegraphics[width=8cm]{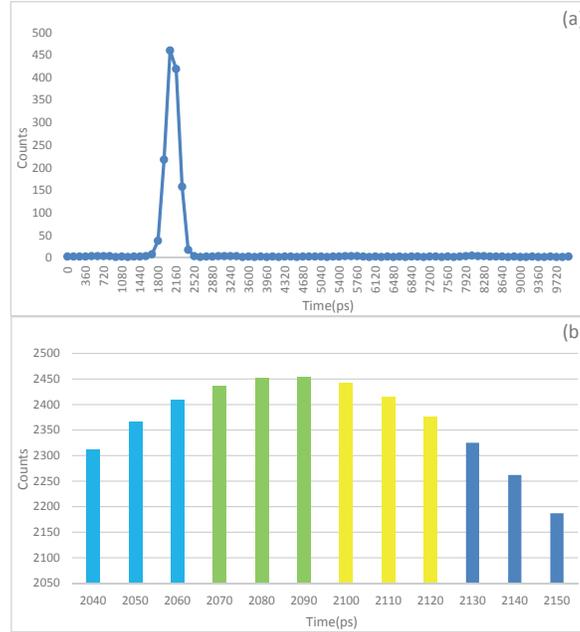}
\caption{The timing response histogram of the detected counts of implementation method I.
a, low-precision synchronization.
The time search range of ${\rm SPD_{1}}$ is [0, 2.5] ns, the time search range of ${\rm SPD_{2}}$ is [2.5, 5.0] ns, the time search range of ${\rm SPD_{3}}$ is [5.0, 7.5] ns, and time search range of ${\rm SPD_{4}}$ is [7.5, 10.0] ns.
b, high-precision synchronization. The time search range of ${\rm SPD_{1}}$ is [2040, 2060] ps, the time search range of ${\rm SPD_{2}}$ is [2070, 2090] ps, the time search range of ${\rm SPD_{3}}$ is [2100, 2120] ps, and time search range of ${\rm SPD_{4}}$ is [2130, 2150] ps. }
\label{figure:Implementation Method I}
\end{figure}

We divide the synchronization process into two round processes.
The first round is low-precision synchronization, and the second round is high-precision synchronization.
We fixed the detector count statistical time to 1ms in low-precision synchronization and 5ms in high-precision synchronization.
The fixed high-precision is 10ps.
In low-precision synchronization, we need to optimize the synchronization precision parameter $t$.
The time $T_1$ taken for synchronization initialization is calculated as follows,

\begin{equation}
T_1 = min\{\lceil \frac{10000}{4 \times t} \rceil \times 1 + \lceil \frac{t}{4 \times 10} \rceil \times 5 \},
\end{equation}
where we have $10 \le t \le 500$.
The optimal $t$ is 120 ps, and the total time $T_1$ spent in the synchronization calibration process of method I is 36 ms.
The synchronization calibration process of method I is shown in Fig. \ref{figure:Implementation Method I}.

\subsection{Method II}

Although the total time spent in the synchronization calibration process of method I is reduced, this method is still not optimal.
Here, we discuss how the optimal synchronization calibration process is implemented.
Same as method I, we divide the synchronization process into two round processes. The difference lies in that we use the dichotomy algorithm with four detectors in the high-precision synchronization process.
In the low-precision synchronization process, The time search range of ${\rm SPD_{1}}$ is [0, 2.5] ns, the time search range of ${\rm SPD_{2}}$ is [2.5, 5.0] ns, the time search range of ${\rm SPD_{3}}$ is [5.0, 7.5] ns, and time search range of ${\rm SPD_{4}}$ is [7.5, 10.0] ns.
After the low-precision synchronization process, we find the time window corresponding to the maximum count is $1.92$ ns, and the corresponding detector is ${\rm SPD_{1}}$.
Therefore, the first time search range of high-precision synchronization is [1.44, 2.88] ns.
Then, we divide this time search range into four equal parts, so that the time window for the first search are $1.62$ ns for ${\rm SPD_{1}}$, $1.81$ ns for ${\rm SPD_{2}}$, $2.00$ ns for ${\rm SPD_{3}}$ and $2.19$ ns for ${\rm SPD_{4}}$, respectively.
The subsequent operation is similar, and the detailed processing and results are shown in the following Fig. \ref{figure:Implementation Method III}.
The total time $T_2$ spent in the synchronization calibration process of method II is 20 ms.

\begin{figure}[h!]
\centering
  \includegraphics[width=8cm]{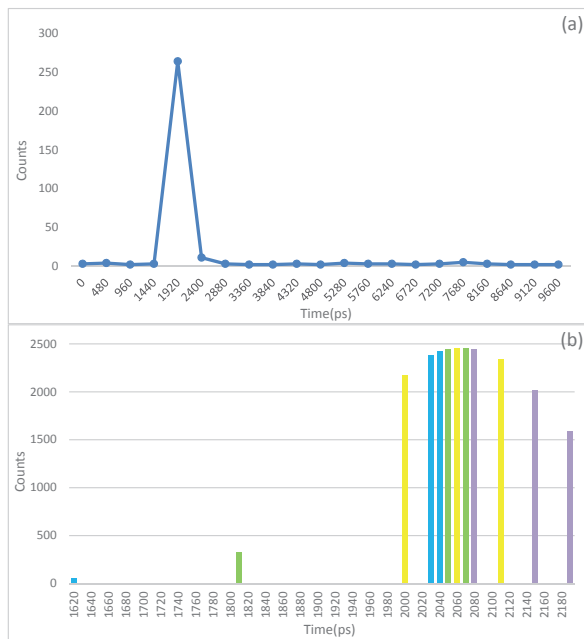}
\caption{The timing response histogram of the detected counts of implementation method III.
a, low-precision synchronization.
The time search range of ${\rm SPD_{1}}$ is [0,2.5] ns, the time search range of ${\rm SPD_{2}}$ is [2.5, 5.0] ns, the time search range of ${\rm SPD_{3}}$ is [5.0, 7.5] ns, and time search range of ${\rm SPD_{4}}$ is [7.5, 10.0] ns.
b, high-precision synchronization.
The time windows for the first round are $1.62$ ns for ${\rm SPD_{1}}$, $1.81$ ns for ${\rm SPD_{2}}$, $2.00$ ns for ${\rm SPD_{3}}$, and $2.19$ ns for ${\rm SPD_{4}}$.
The time windows for the second round are $2.03$ ns for ${\rm SPD_{1}}$, $2.07$ ns for ${\rm SPD_{2}}$, $2.11$ ns for ${\rm SPD_{3}}$, and $2.15$ ns for ${\rm SPD_{4}}$.
The time windows for the third round are $2.04$ ns for ${\rm SPD_{1}}$, $2.05$ ns for ${\rm SPD_{2}}$, $2.06$ ns for ${\rm SPD_{3}}$, and $2.08$ ns for ${\rm SPD_{4}}$.
}
\label{figure:Implementation Method III}
\end{figure}

\section{Discussion and Conclusion}

In Ref. ~\cite{liang2012fully}, the time required for the synchronization calibration process is about one minute.
In Ref. ~\cite{rosenberg2009practical}, the time required for the synchronization calibration process is typically 1-10 seconds.
In comparison, the parallel synchronization method in this paper can only require tens of milliseconds. We remark that the synchronization time required by our parallel rules is at most $1/N$ of the previous one \cite{liang2012fully,rosenberg2009practical} even under the same conditions, where $N$ is the number of detectors.
For the QKD system based on the Large-alphabet protocol~\cite{ali2007large,lee2016high}, the number of detectors used at the receiver is generally not less than 8.
The time of synchronization calibration process can be reduced even more.
During the synchronization process, Eve can shift the arrival time of each pulse sent from Alice by employing a variable optical delay line.
This type of attack causes the signal time window of multiple detectors to shift forward or backward as a whole.
Under the condition that the time efficiency curves of multiple detectors are consistent or close to each other, such an attack will only reduce the detection efficiency of the QKD system without causing the detector time-dependent efficiency mismatch in our new synchronization scheme.

In summary, an efficient synchronization method without security loophole is proposed.
This synchronization method adopts the parallel search method, and divides the time search range of the synchronization process depending to the number of detectors at the receiver, and then all the time search simultaneously in their respective search ranges.
Through this parallel search method, the synchronization initialization time can be greatly reduced without security loophole.
As the number of detectors increases, the time of bit synchronization is significantly decreased.
Most of the current QKD systems use a multi-detector scheme, the application scenario of this synchronization method is expected to be very wide.

\section{Acknowledgments}
We gratefully acknowledge support from the National Natural Science Foundation of China under Grant No. 61801420, the Fundamental Research Funds for the Central Universities.

\bibliographystyle{apsrev}



\end{document}